\documentstyle[12pt,fleqn]{article}
\begin{document}

\textheight 8.5in
\topmargin -.25in
\oddsidemargin 0in
\evensidemargin 0in

\jot=4mm
\textwidth=18truecm
\baselineskip=24pt    
\parindent=20pt
\headheight=12pt
\footskip=48pt
\hoffset=-0.9cm
\oddsidemargin=1cm %
\hsize=15.5truecm  %
\setlength{\unitlength}{.1cm}

\def\beqa{\begin{eqnarray}}
\def\eeqa{\end{eqnarray}}
\def\beq{\begin{equation}}
\def\eeq{\end{equation}}
\def\beqal{\begin{eqnarray}\label}
\def\beql{\begin{equation}\label}

\def\half{\frac{1}{2}}
\def\R{\relax{\rm I\kern-.18em R}}
\def\uno{{1 \kern-.28em {\rm l}}}
\def\Ds{\ {\big / \kern-.70em D}}
\def\C{{\kern .1em {\raise .47ex \hbox {$\scriptscriptstyle |$}}
\kern -.5em {\rm C}}}
\def\Z{{Z \kern-.45em Z}}
\def\Q{{\kern .1em {\raise .47ex \hbox{$\scripscriptstyle |$}}
\kern -.35em {\rm Q}}}
\def\de{{\rm d}}
\def\Tr{{\rm Tr}}
\def\tr{{\rm tr}}
\def\psb{\bar{\psi}}
\def\chb{\bar{\chi}}
\def\lb{\bar{\lambda}}
\def\epsb{\bar{\varepsilon}}
\def\sb{\bar{\sigma}}
\def\ds{\partial\!\!\!\!\!\!\;\:/}
\def\p{\partial}

\def\al{\alpha}
\def\b{\beta}
\def\g{\gamma}
\def\d{\delta}
\def\eps{\varepsilon}
\def\z{\zeta}
\def\h{\eta}
\def\th{\theta}
\def\k{\kappa}
\def\l{\lambda}
\def\m{\mu}
\def\n{\nu}
\def\x{\xi}
\def\r{\rho}
\def\s{\sigma}
\def\t{\tau}
\def\ph{\phi}
\def\ch{\chi}
\def\ps{\psi}
\def\om{\omega}
\def\G{\Gamma}
\def\D{\Delta}
\def\Th{\Theta}
\def\L{\Lambda}
\def\X{\Xi}
\def\P{\Pi}
\def\S{\Sigma}
\def\Ph{\Phi}
\def\Ps{\Psi}
\def\O{\Omega}

\def\jmp{{\it J. Math. Phys.}\ }
\def\prd{{\it Phys. Rev. {\bf D}}\ }
\def\pre{{\it Phys. Rev.}\ }
\def\jp{{\it J. Phys.}\ }
\def\prl{{\it Phys. Rev. lett.}\ }
\def\pl{{\it Phys. Lett.}\ }
\def\npb{{\it Nucl. Phys. {\bf B}}\ }
\def\mpl{{\it Mod. Phys. Lett.}\ }
\def\ap{{\it Ann. Phys.}\ }
\def\ijmp{{\it Int. J. Mod. Phys.}\ }
\def\cmp{{\it Comm. Math. Phys.}\ }
\def\cqg{{\it Class. Quant. Grav.}\ }
\def\spj{{\it Sov. Phys. JETP}\ }
\def\spjl{{\it Sov. Phys. JETP lett.}\ }
\def\prs{{\it Proc. R. Soc.}\ }
\def\grg{{\it Gen. Rel. Grav.}\ }
\def\nat{{\it Nature}\ }
\def\apj{{\it Astrophys. J.}\ }
\def\aaa{{\it Astron. Astrophys.}\ }
\def\ncim{{\it Nuovo Cim.}\ }
\def\ptp{{\it Prog. Theor. Phys.}\ }
\def\aip{{\it Adv. Phys.}\ }
\def\jpamg{{\it J. Phys. A: Math. Gen.}\ }
\def\mnras{{\it Mon. Not. R. Astr. Soc.}\ }
\def\prep{{\it Phys. Rep.}\ }
\def\ncb{{\it Il Nuovo Cimento ``B''}}
\def\ssr{{\it Space Sci. Rev.}\ }
\def\pasp{{\it Pub. A. S. P.}\ }
\def\araa{{\it Ann. Rev. Astr. Ap.}\ }
\def\asr{{\it Adv. Space Res.}\ }
\def\etal{{\it et al.}\ }
\def\ie{{\it i.e. }}
\def\eg{{\it e.g. }}

\begin{titlepage}
\begin{flushright}
ROM2F-95-34\\
\today
\end{flushright}
\vspace{1.0cm}
\begin{center}

{\large \bf Fermionic zero-modes around string solitons}\\

\vspace{0.5cm}

{\bf Diego Bellisai}
\footnote{
E-mail: {\it bellisai@roma2.infn.it, bellisai@roma1.infn.it}}

\vspace{0.1cm}

{\sl Dipartimento di Fisica, Universit\`{a} di Roma I ``La Sapienza" \\
Piazzale Aldo Moro 5, 00185 Rome, Italy, \\
and\\
I.N.F.N. \ - \ Sezione di Roma II ``Tor Vergata",
00133 Rome, Italy}

\vspace{1.0cm}
\end{center}
\abstract{\it In presence of string solitons, index theorems for the
generalised Dirac operators have to be revisited. We show that in
supersymmetric configurations the fermionic operators decouple, so that
there are no mixing effects between different fermions in the index
theorems. We extend the index theorems in presence of
torsion to the generic case of manifolds with
boundary, which naturally appear in string solutions
and apply this result to the soliton solution by Callan, Harvey
and Strominger.}
\vfill
\end{titlepage}
\newpage
The existence of solitons
means that a full non-perturbative theory may have a much richer structure
than it is apparent in perturbation theory: since it is commonly believed
that pointlike
quantum field theories are not adequate to unify all the four
fundamental forces, including gravity, and that they should be substituted
by string theory \cite{GSW}, it is important to study soliton solutions
in this context.
In the last few years much attention has been
paid to the existence
of these solutions after the paper by Strominger \cite{STR}, who showed
that supergravity in ten dimensions coupled to super-Yang-Mills, which
 is the field theory limit of the heterotic string, has as solitonic
solution the heterotic fivebrane (a five-dimensional extended object),
already
conjectured by Duff two years before \cite{DUFF}. This fivebrane
is everywhere
nonsingular and carries a topological charge,
just as usual 't Hooft instantons in four dimensions.
After this solution many others were found
 \cite{KHURIREP},
some of which
were exact to all orders in $\al^{\prime}$ the inverse
string tension and which, in
certain limits, corresponded to exact conformal field theories
given by a Wess-Zumino-Witten model times a one-dimensional
Feigin-Fuchs Coulomb gas \cite{CHS}. The common feature is given by the
presence of a Yang-Mills instanton in the four directions transverse to the
fivebrane.
Just as in the 4-dimensional instantonic case, we are interested in the
 number and
structure of the fermionic zero-modes, whose presence could drive
the breaking
of chiral symmetries and/or supersymmetry, through
the formation of gaugino \cite{NILLES,FERGIRNIL} or
gravitino condensates \cite{WIT,KONMAGN,HAW}.
A first step
therefore is to apply the index theorems to the fermionic kinetic operators
to compute the number of zero-modes these solitonic solutions have.
A novel feature is the presence of the 3-form $H$
(the field-strength of the two form $B$ plus Chern-Simons terms)
in the supergravity Lagrangian, which plays the role of a torsion and
which may affect the index formulae.
The importance played by the antisymmetric tensor in superstring theory and
in compactification has been greatly stressed by Rohm and Witten \cite{ROHM}
and by Strominger \cite{STROMTOR};
up to now the Dirac index formula for manifolds with torsion \cite{YAJ}
has been known only in the case of compact manifolds without boundary and
studied in a totally different context.
In our case we want to get in touch with the kind of Dirac
operators arising from the compactification of superstring inspired
supergravity Lagrangians.
We propose an extension of this formula to the case of manifolds with
boundary and apply it to
the solution of Callan, Harvey and Strominger (CHS)
\cite{CHS} and derive the exact form of the fermionic operators; from
there we will infer the number of zero-modes for gaugino and dilatino.
The formulae we derive have a wide applicability, since we have found that
for supersymmetric solutions there is a decoupling among the fermionic
operators which makes possible to neglect mixing effects between the
different fermion indices, although singularities in the string solution
may require a subtle extension.
In Section 1 we briefly sketch the conditions for having
supersymmetric backgrounds exact to all orders in $\al^{\prime}$.
In Section 2 we explicitly calculate the
fermionic equations of motion in supersymmetric
backgrounds and show their factorisation. In
Section 3 we review the formulation of Dirac index theorem in presence of
torsion and introduce our generalisation to manifolds with boundaries.
Finally in Section 4 we apply the
results of Sect. 3 to a specific soliton solutions and calculate the number
of zero-modes of the fermionic operators.
\section{Supersymmetric backgrounds}
First of all, let us introduce the model we want to study:
we start from the $D=10$ $N=1$ supergravity
and super Yang-Mills action
\cite{BDR} which coincides with the leading terms coming
from the heterotic string theory. The bosonic part of the action is:
\beq
S_{B}=-\half\int\de^{10}x\sqrt{g}e^{-2\Phi}\left(R-4(\nabla\Phi)^{2}+
\frac{1}{3}H_{MNP}H^{MNP}+\frac{\al^{\prime}}{30}\Tr F_{MN}F^{MN}\right),
\eeq
where $H$ is related to the antisymmetric tensor $B$ by the following
relation:
\beq
H=\de B+\al^{\prime}\left(\S_{3}^{L}(\O_{-})-\frac{1}{30}\S_{3}^{YM}(A)
\right),
\eeq
where $\S_{3}$ stands for the Chern-Simons three-form, so that
we obtain modified Bianchi identities for $H$:
\beq
\de H=\al^{\prime}\left(\tr\ R(\O_{-})\wedge R(\O_{-})
-\frac{1}{30}\Tr\ F\wedge F\right).
\eeq
The trace $\Tr$ is over the adjoint representation of the gauge group, and
the curvature $R$ is built with
$\O_{-M}{}^{AB}=(\om{_M}{}^{AB}-H{_M}{}^{AB})$, a generalised spin
connection with torsion. Instead of solving the equations of motions for
this action, it is more convenient
and rewarding to look for bosonic backgrounds
annihilated by some of the $N=1$ supersymmetry transformations, since only
the vacuum is annihilated by all of them.
The supersymmetry transformations of the fermionic fields are:
\beqa
\d\chi&=&-\frac{1}{4}F_{MN}\G^{MN}\eps,\nonumber\\
\d\l&=&-\frac{1}{4}\left(\G^{M}\p_{M}\Phi-\frac{1}{6}H_{MNP}\G^{MNP}\right)
\eps,\\
\d\psi_{M}&=&\left(\p_{M}+\frac{1}{4}\O_{-M}{}^{AB}\G_{AB}\right)\eps.
\nonumber
\eeqa
For simplicity we take
all fields indipendent on the six Minkowskian coordinates of the fivebrane.
In addition we must take into
account that the supersymmetry spinor in ten dimensions belongs to the
representation {\bf 16} of $SO(1,9)$ and that under $SO(1,5)\times SO(4)$
it breaks in ${\bf 16}=({\bf 4},{\bf 2_{+}})\oplus({\bf 4^{\ast}},
{\bf 2_{-}})$ which we call $\eps_{\pm}$.
The $D=4$ bosonic background
we are interested in annihilates $\eps_{+}$: then we find that the most
general solution is given by the configuration:
\beq
F=\pm\ast F,\quad\quad H=\pm\ast\de\Phi,\quad\quad R(\O)=\pm\ast R(\O),
\eeq
where $\O=(\om+H)$ stands for a generalised spin connection with torsion.
If in addition we require a solution exact to all orders in $\al^{\prime}$,
it is possible to show that we are forced to identify the generalised spin
connection with the gauge field through the procedure known as standard
embedding.
It is then possible to
distinguish two cases; either the dilaton and the three-form $H$ are taken
to be zero and the metric is self-dual, corresponding to a gravitational
instanton\cite{BFRM,KKL},
or the dilaton and $H$ are non-trivial while the
metric is conformally flat \cite{REGGE,KHURI,CHS}
leaving us with a
$SU(2)$ selfdual connection $\O_{-M}{}^{AB}$, while $\O_{+M}{}^{AB}$ is
anti-selfdual.
To make the gaugino variation vanish it suffices to take the gauge field to
be a 't Hooft instanton;
$\d\chi$ vanishes if $\eps=({\bf 4},{\bf 2_{+}})$
and
$F_{\m\n}=\tilde{F}_{\m\n}$,
where Greek indices run from 1 to 4, the coordinates transverse to the
fivebrane,
We take the field strength of
the antisymmetric tensor to be
$H_{\m\n\r}=-\sqrt{g}\eps_{\m\n\r}{}^{\s}\p_{\s}\Phi$ to annihilate the
negative chirality dilatino $\l$ transformation and
the four dimensional metric conformally flat
$g_{\m\n}=e^{2\Phi}\d_{\m\n}$.
Taking a constant positive chirality supersymmetry spinor we also
annihilate the positive chirality gravitino variation; finally
we make the identification
$R_{\m\n}(\O_{+})^{ab}=\half\bar{\eta}^{I,ab}F^{I}_{\m\n}$, where
$\bar{\eta}^{I}_{ab}$ are the 't Hooft symbols, in order to annihilate the
$\al^{\prime}$ correction to the Bianchi identities and, more in general,
all the other corrections in $\al^{\prime}$.
As already alluded to, this procedure has a close analogy with the standard
embedding in Calabi-Yau compactification of the heterotic string
\cite{CHSW,STROMTOR}
used to get an $N=1$ $D=4$ supersymmetric background from
an $N=1$ $D=10$ string theory\footnote{In that case the embedding is
between the $SU(3)$ spin connection corresponding to the internal
manifold holonomy
group and the connection of an $SU(3)$ subgroup of the gauge group, that
is $SO(32)$ or $E_{8}\times E_{8}$. In the heterotic string case this
embedding is
in the first $E_{8}$ corresponding to the gauge group which will give rise
to the low energy phenomenology, while the second $E_{8}$
corresponds to the
so-called {\it hidden sector} which interacts only gravitationally with
the particles
belonging to the first sector.}. In string soliton solutions,
the embedding is
between an $SU(2)$ connection with torsion of the four dimensional manifold
transverse to the fivebrane and an $SU(2)$ subgroup of the second $E_{8}$.
The Bianchi identities become $e^{-2\Phi}\Box e^{2\Phi}=0$,
where the d'Alembertian is intended to be evaluated in flat
Euclidean space.
According to the number of coordinates $\Phi$ is supposed to depend on, one
has
\begin{enumerate}
\item five-branes, characterised by a dilaton field
$e^{2\Phi}=e^{2\Phi_{0}}+\sum_{i}\frac{Q_{i}}{(x-x_{i})^{2}}$,
considered by CHS \cite{CHS};
\item monopoles, with
$e^{2\Phi}=e^{2\Phi_{0}}+\sum_{i}\frac{m_{i}}{|\vec{x}-\vec{x}_{i}|}$ found
in \cite{KHURILET,KHURI,GAUNT};
this solution corresponds in four dimensional Minkowskian space
to pointlike topological defects localised in the
points $\vec{x}_{i}$;
\item strings with
$e^{2\Phi}=e^{2\Phi_{0}}+\sum_{i}q_{i}\log|z-z_{i}|^{2}$,
where $z=x_{1}+ix_{2}$, corresponds to
one-dimensional defects \cite{STRING};
\item domain walls with
$e^{2\Phi}=e^{2\Phi_{0}}+\sum_{i}c_{i}(t-t_{i})$, corresponding to
two-dimensional topological defects \cite{STRING}.
\end{enumerate}
Under Buscher's duality \cite{BUSCH} these solutions are related to purely
gravitational backgrounds with zero torsion but a non-trivial dilaton
field.
In the following we will concentrate on solution 1.
As we can see, this solution breaks two of the four supersymmetries,
related
to the positive chirality supersymmetry parameter.
\section{Fermionic equations of motion}
We are now ready to discuss the
fermionic equations of motion arising from the ten-dimensional $N=1$
supergravity plus super-Yang-Mills Lagrangian
which represents the pointlike limit
of the heterotic string theory. This Lagrangian has the form \cite{BDR}:
\beqa
e^{-1}L_{tot}&=&e^{-2\Phi}\left\{-\half
R-\frac{1}{6}H_{MNP}H^{MNP}+
2(\p_{M}\Phi)^{2}-\frac{1}{4}F_{MN}^{\al}F^{MN\al}-\right.\cr\cr
&-&\left.\half\chb^{\al}\G^{M}{\cal D}_{M}\chi^{\al}-
\half\psb_{M}\G^{MNP}D_{N}\psi_{P}-\psi_{M}\G^{M}\p^{N}\Phi\psi_{N}
-\lb\G^{M}D_{M}\l-\right.\cr\cr
&-&\left.\half\psb_{M}\G^{N}
\p_{N}\Phi\G^{M}\l+\frac{1}{24}H_{RST}\left[\psb_{M}
\G^{MRSTN}\psi_{N}+
6\psb^{R}\G^{S}\psi^{T}+
\right.\right.\\ \cr
&+&\left.\left.2\psb_{M}(\G^{MRST}-
3g^{MT}\G^{RS})\l\right]-
\frac{1}{4}\chb^{\al}(\G^{MNP}F_{NP}^{\al}+
2F^{M}{}_{P}{}^{\al}\G^{P})\psi_{M}-\right.\cr\cr
&-&\left.\frac{1}{4}
\chb^{\al}\G^{NP}F_{NP}^{\al}\l+\frac{1}{24}
\Tr(\chb\G^{MNP}\chi)H_{MNP}\right\}\nonumber.
\eeqa
where $\al$ is an $E_{8}$ index and
where we have redefined the gravitino $\psi_{M}$ to make diagonal
the kinetic derivative terms of the fermions; our gravitino is
related to that of Bergshoeff and de Roo by:
\beq
\psi_{M}=\psi_{M}^{(BdR)}+\frac{\sqrt{2}}{4}\G_{M}\l^{(BdR)}.
\eeq
Moreover to agree with \cite{CHS}, we have made some other
field redefinitions:
\beq
H_{MNP}=\frac{3}{\sqrt{2}}H_{MNP}^{(BdR)},\qquad\l=\frac{1}
{\sqrt{2}}\l^{(BdR)},\qquad
\Phi=\frac{3}{2}\log\phi^{(BdR)}.
\eeq
{}From this Lagrangian it is now easy to derive the fermionic equations of
motion, taking into account that all the fermions are represented by
Majorana-Weyl spinors: the equation of motion of the gravitino is
\beqa
e^{-1}\frac{\d S}{\d\psb_{M}}&=&e^{-2\Phi}\left\{-\G^{MNP}\left(
D_{N}-\p_{N}\Phi\right)\psi_{P}+
\frac{1}{12}H_{RST}\G^{MRSTN}\psi_{N}-\right.\nonumber\\
&-&\left.\half\left(\G^{N}\p_{N}\Phi+\frac{1}{6}
H_{RST}\G^{RST}\right)\G^{M}\l-2\G^{M}\p^{N}\Phi\psi_{N}\right.\\
&+&\left.
\half H^{MNP}\G_{N}\psi_{P}-\frac{1}{4}(\G^{MNP}F_{NP}^{\al}-
2F^{M}{}_{P}{}^{\al}\G^{P})\chi^{\al}\right\}=0\nonumber;
\eeqa
the equation for the dilatino is:
\beqa
e^{-1}\frac{\d S}{\d\lb}&=&e^{-2\Phi}\left\{-2\G^{M}\left(
D_{M}-\p_{M}\Phi\right)\l
+\frac{1}{4}
\G^{NP}F^{\al}_{NP}\chi^{\al}-\right.\\
&-&\left.
\half\G^{M}\left(\G^{N}\p_{N}\Phi-
\frac{1}{6}
\G^{RST}H_{RST}\right)\psi_{M}\right\}=0\nonumber,
\eeqa
and the equation for the gaugino is:
\beqa
e^{-1}\frac{\d S}{\d\chb^{\al}}&=&e^{-2\Phi}
\left\{-\G^{M}\left(
{\cal D}_{M}-\p_{M}\Phi\right)\chi+
\frac{1}{12}\G^{MNP}H_{MNP}\chi^{\al}-\right.\cr\cr
&-&\left.\frac{1}{4}(\G^{MNP}F_{NP}^{\al}+
2F^{M}{}_{P}{}^{\al}\G^{P})\psi_{M}-\frac{1}{4}
\G^{NP}F_{NP}^{\al}\l\right\}=0.
\eeqa
As we have shown before, supersymmetric solitons are characterised
by background fields completely independent of the
coordinates of the six-dimensional Minkowskian manifold
and depending on some of the remaining coordinates
of the four-dimensional curved manifold.
To obtain the four-dimensional equations of motion we simply
have to neglect the components of the
background fields with
indices on the six-dimensional flat Minkowski manifold swept out by the
five-brane.
Supersymmetric solitons half of the supersymmetries, since
they annihilate the positive chirality four-dimensional
supersymmetry parameter; thus, since we are interested in the
possible presence of fermionic zero-modes, we have to study
the equations of motion for negative chirality gravitino and gaugino and
for positive chirality dilatino.
Taking into account that
\beqa
& &(\g^{\m\n\r}F^{I}_{\n\r}-2F^{\m}{}_{\r}{}^{I}\g^{\r})\eps_{\pm}=
(\sqrt{g}\eps^{\m\n\r\s}\g_{5}\g_{\s}F^{I}_{\n\r}-2F^{\m}{}_{\r}{}^{I}
\g^{\r})
\eps_{\pm}=\cr
& &=-2(\tilde{F}^{\m}{}_{\r}{}^{I}\g_{5}-F^{\m}{}_{\r}{}^{I})
\g^{\r}\eps_{\pm}=
-2F^{\m}{}_{\r}{}^{I}\g^{\r}(\uno+\g_{5})\eps_{\pm},
\eeqa
where $I$ is an $SU(2)$ index, and $\m,\n,\r,\s$ are curved 4-dimensional
indices,
we obtain the following equations:
\beq
-\g^{\m\n\r}(D_{\n}-\p_{\n}\Phi)\psi_{\r}^{-}-
2\g^{\m}\p^{\n}\Phi\psi_{\n}^{-}
+\half H^{\m\n\r}\g_{\n}\psi_{\r}^{-}=0,
\eeq
\beq
-2\g^{\m}(D_{\m}-\p_{\m}\Phi)\l_{+}-\g^{\m}\g^{\n}\p_{\n}\Phi
\psi_{\m}^{-}+\frac{1}{4}F^{I}_{\m\n}\g^{\m\n}\chi^{I}_{-}=0,
\eeq
\beq
-\g^{\m}({\cal D}_{\m}-\p_{\m}\Phi)\chi^{I}_{-}+\frac{1}{12}H_{\m\n\r}
\g^{\m\n\r}\chi^{I}_{-}-F^{\m}{}_{\r}{}^{I}\g^{\r}\psi_{\m}^{-}=0.
\eeq
As we can see,
the generalised Dirac operator acting on
fermions is represented by a triangular matrix, so that its determinant
is given by the product of the
determinants of the fermionic kinetic operators.
We are then authorised to calculate separately
the index theorems for each of these fermion operators,
without worrying about mixing effects between
different fermions\cite{YAJMOD}.
The equations relevant to this calculation are then:
\beqa
& &\g^{\m\n\r}(D_{\n}-\p_{\n}\Phi)\psi_{\r}^{-}+
2\g^{\m}\p_{\n}\Phi\psi_{\n}^{-}
-\half H^{\m\n\r}\g_{\n}\psi_{\r}^{-}=0,\nonumber \\
& &\g^{\m}(D_{\m}-\p_{\m}\Phi)\l_{+}=0,\\
& &\g^{\m}({\cal D}_{\m}-\p_{\m}\Phi)\chi^{I}_{-}-\frac{1}{12}H_{\m\n\r}
\g^{\m\n\r}\chi^{I}_{-}=0.\nonumber
\eeqa
Defining new dilatino and gaugino fields $\hat{\l}=e^{-\Phi}\l$,
$\hat{\chi}^{I}=e^{-\Phi}\chi^{I}$ we get the simple equations
\beqa
& &\g^{\m}D_{\m}\hat{\l}_{+}=0,\\
& &\g^{\m}{\cal D}_{\m}\hat{\chi}^{a}_{-}-\frac{1}{12}H_{\m\n\r}
\g^{\m\n\r}\hat{\chi}^{a}_{-}=0,\nonumber
\eeqa
which are well suited to be investigated through the index theorems.
Regarding the gravitino equation, we must get
rid of the unwanted terms containing the dilaton field; in general
this will not be possible. We will however show that for
string solitons this is possible.
\section{Index theorems in presence of torsion}
Index theorems represent a fundamental tool in studying the structure
of differential elliptic operators like the exterior derivative on forms,
the Dirac operators on curved backgrounds or coupled to non-trivial gauge
connections, and finally the Rarita-Schwinger operator for the
gravitino.
They give us information about the existence of zero-modes of these
operators and
on their number or, better, on certain algebraic sums of these modes.
These theorems are well known
for manifolds without torsion, that is in the case
in which the affine connections $\G^{\l}{}_{\m\n}$ are symmetric in their
lower indices, and thus coincide with the Christoffel connection.
In a generic manifold with torsion, the relation between the affine
connection with torsion
$\G^{\l}{}_{\m\n}$ and the spin connection $\O_{\m}{}^{ab}$
is still given by
the metricity condition:
\beq
D_{\m}e^{a}_{\n}=\p_{\m}e^{a}_{\n}+\O_{\m}{}^{a}{}_{b}e^{b}_{\n}-
\G^{\l}{}_{\n\m}e^{a}_{\l}=0,
\eeq
from which we have
\beq
\O_{\m}{}^{ab}=\om_{\m}{}^{ab}+e^{a}_{\n}e^{b}_{\r}K^{\n\r}
{}_{\m},\quad\quad
\G^{\l}{}_{\m\n}=\g^{\l}{}_{\m\n}+K^{\l}{}_{\m\n},
\eeq
where $\g^{\l}{}_{\m\n}$ are the Christoffel
symbols, $K^{\l}{}_{\m\n}$ is the contortion tensor
\beq
K^{\l}{}_{\m\n}=\half(T_{\m\n}{}^{\l}+T^{\l}{}_{\m\n}+T^{\l}{}_{\n\m}),
\eeq
defined in terms of the torsion tensor
$T^{\l}{}_{\m\n}=\G^{\l}{}_{\m\n}-\G^{\l}{}_{\n\m}$,
and $\om_{\m}{}^{ab}$ is the
Levi-Civita spin connection;
the commutation relation between the covariant derivatives is:
\beq\label{commut}
[D_{\m},D_{\n}]=\half R^{ab}{}_{\m\n}\S_{ab}+T^{\l}{}_{\m\n}D_{\l},
\eeq
where $\S_{ab}=\frac{1}{4}[\g_{a},\g_{b}]$
are the generators of the Lorentz transformations.
The derivation of the chiral anomaly in the Riemann-Cartan space, following
the articles by Yajima and Kimura \cite{YAJ}, is obtained through
De Witt's heat kernel method \cite{DEW} applied to the Dirac operator with
a torsionful spin connection. The fundamental step is to
introduce a new spin connection which cancels the linear term in the
covariant derivative in (\ref{commut}) to calculate all the quantities
depending on the spin connection with the new spin
connection with torsion.
After some manipulations it turns out that the chiral anomaly is given by
$A(x)=\frac{1}{16\pi^{2}}\Tr\g_{5}[a_{2}]$,
where $[a_{2}]$ is the so-called second De Witt-Minakshisundaram-Seeley
coefficient:
\beqa
[a_{2}]&=&\frac{1}{12}Y^{\m\n}Y_{\m\n}+\frac{1}{180}
(R^{\m\n\r\s}R_{\m\n\r\s}-R^{\m\n}R_{\m\n})+\cr\cr
&+&\frac{1}{6}\stackrel{\sim}{\Box}\left(Z-\frac{1}{5}R\right)+
\half\left(Z-\frac{1}{6}R\right)^{2}.
\eeqa
The tensor $Y_{\m\n}$ is given by
\beq
Y_{\m\n}=\frac{1}{4}\tilde{R}^{ab}{}_{\m\n}\g_{ab}-F_{\m\n},
\eeq
where the tilde stands for objects calculated with the new spin connection
which has three times the original torsion,
and, in case of totally antisymmetric torsion,
\beq
Z=\frac{1}{4}R+\frac{1}{8}S^{\m}S_{\m}+\frac{1}{4}\g_{5}\nabla_{\m}S^{\m}-
\half\g^{\m\n}F_{\m\n},
\eeq
where $S_{\m}=-\half\eps_{\m\n\r\s}K^{\n\r\s}$, $\eps_{\m\n\r\s}$
is a covariantly constant tensorial density and $\nabla_{\m}$ means
covariant differentiation with respect to the Levi-Civita connection.
Finally, using the reduction formulae of the $\g_{ab}$ matrices
Yajima and Kimura obtained:
\beqa
A(x)&=&\frac{1}{16\pi^{2}}\left\{\eps_{abcd}\left(-\frac{{\rm dim}\ V}{48}
\tilde{R}^{ab\m\n}\tilde{R}^{cd}{}_{\m\n}+\half\tr(F^{ab}F^{cd})\right)+
\right.\cr\cr
&+&\left.\frac{{\rm dim}\ V}{6}\Box(\nabla_{\m}S^{\m})+
\frac{{\rm dim}\ V}{12}R\nabla_{\m}
S^{\m}+
\frac{{\rm dim}\ V}{8}S^{\n}S_{\n}\nabla_{\m}S^{\m}\right\}
\eeqa
where ${\rm dim}\ V$ is the dimension of
the representation $V$ of the gauge group $G$
the Dirac fermion belongs to (if $G=SU(2)$, then ${\rm dim}\ V=2t+1$).
This result is valid for manifold without boundary.
In the case of supergravity
where the role of torsion is played by
the totally antisymmetric field-strength tensor $H_{\m\n\r}$ which
satisfies the Bianchi identities:
\beq
\eps^{\m\n\r\s}\nabla_{\m}H_{\n\r\s}=0,
\eeq
equivalent to $\nabla_{\m}S^{\m}=0$,
we obtain a powerful simplification
which allows us to extend the index theorem to manifolds with boundaries.
The volume contribution to the anomaly is now simply:
\beq
A(x)=\frac{1}{16\pi^{2}}\eps_{abcd}\left\{-\frac{{\rm dim}\ V}{48}
\tilde{R}^{ab\m\n}\tilde{R}^{cd}{}_{\m\n}+\half\tr(F^{ab}F^{cd})\right\},
\eeq
which translated in terms of differential forms and characteristic
polynomials becomes:
\beq
{\rm ind}\Ds(M,V)=\frac{2t+1}{192\pi^{2}}\int_{M}\tr(\tilde{R}(\tilde{\O})
\wedge
\tilde{R}(\tilde{\O}))-\frac{1}{8\pi^{2}}\int_{M}\Tr(F\wedge F),
\eeq
where $\tilde{\O}=\om-3K$, $M$ is a four dimensional manifold and
where the curvature two-form $\tilde{R}$ is calculated from the
connection with
{\it minus} three times the torsion, due to the exchange symmetry
of the torsionful Riemann curvature tensor:
\beq
\tilde{R}_{\m\n\r\s}(\om+K)=\tilde{R}_{\r\s\m\n}(\om-K),
\eeq
when the torsion satisfies the Bianchi identities \cite{MAV}.
Exploiting the properties of the invariant polynomials, the
straightforward generalisation we propose is:
\beqa
{\rm ind}\Ds_{V}(M,\p M)&=&\frac{{\rm dim}\ V}{192\pi^{2}}
\left[\int_{M}\tr\tilde{R}(\tilde{\O})
\wedge\tilde{R}(\tilde{\O})
-\int_{\p M}
\tr\ \tilde{\theta}\wedge\tilde{R}(\tilde{\O})\right]-\nonumber\\
&-&\frac{1}{8\pi^{2}}\int_{M}\Tr_{V}(F\wedge F)
-\half[\eta_{D}(\p M)+h_{D}(\p M)]
\eeqa
where the Atiyah-Patodi-Singer (APS)
$\eta$--invariant \cite{APS} is evaluated by solving the
three-dimensional
Dirac equation with spin connection ($\om+K$) on the boundary of
the manifold and the second fundamental form is:
\beq
\tilde{\theta}^{a}{}_{b}=(\tilde{\O})^{a}{}_{b}-(\tilde{\O})^{a}_{(0)b}.
\eeq
If the manifold has more than one boundary, the APS invariant
is given by the
algebraic sum of the invariants
of each boundary, with sign + or -- depending
on the orientation of the boundaries.
The ``0" index means that $\om_{0}$ should be evaluated on a manifold
which admits a product metric on the boundary and that coincides with $\om$
in the bulk, while $K_{0}$ is simply equal to $K$ and so the $K$ contribution
disappears from the definition of the second fundamental form.
\section{Fermionic zero-modes around CHS solitons}
Let us now go to our specific case, the CHS one-instanton solution:
the instanton
gauge field in the language of differential forms reads
$A^{I}\equiv A^{I}_{\m}\de x^{\m}=\frac{2\r^{2}\bar{\s}^{I}}
{\r^{2}+r^{2}}$
where the $\bar{\s}^{I}$ are the $SU(2)$ right-invariant 1-forms.
The standard embedding between the gauge connection and the spin connection
translates into the relation $\O_{+}{}^{ab}=\half\bar{\eta}^{I}_{ab}A^{I}$
where we have chosen the orientation $\eps_{1230}=+1$.
The vierbein is taken to be
$e^{a}=e^{\Phi}(\de r, r\bar{\s}^{1}, r\bar{\s}^{2}, r\bar{\s}^{3})$,
while the Levi-Civita spin connection components are:
\beq
\om^{i0}=(1+r\Phi^{\prime})\sb^{i},\quad\quad
\om^{jk}=\eps^{jki}\sb^{i},
\eeq
where $\Phi^{\prime}=\frac{\de\Phi}{\de r}$.
In the chosen frame, the 1-form $H^{ab}=H_{\m}{}^{ab}\de x^{\m}$
has components
$H^{i0}=0$ and $H^{jk}=-r\Phi^{\prime}\eps^{jki}\sb^{i}$
and the dilaton field is simply given by
$e^{2\Phi}=e^{2\Phi_{0}}\left(1+\frac{\r^{2}}{r^{2}}\right)$,
where we have placed the instanton in the origin
so that:
\beqa
\O_{+}{}^{i0}&=&(1+r\Phi^{\prime})\sb^{i},\;\;\;\;\;
\O_{+}{}^{jk}=(1-r\Phi^{\prime})\eps^{jki}\sb^{i},\cr
\O_{-}{}^{i0}&=&(1+r\Phi^{\prime})\sb^{i},\;\;\;\;\;
\O_{-}{}^{i0}=(1+r\Phi^{\prime})\eps^{jki}\sb^{i}.
\eeqa
Since the gauge instanton lives in an $SU(2)$ subgroup of the
$E_{8}\times E_{8}$ gauge group, we have to decompose the adjoint
representation of the latter
in terms of the representations of the $SU(2)$ the instanton
lives in. By decomposing the second $E_{8}$ with respect to
its maximal subgroup $E_{7}\times SU(2)$, the adjoint {\bf 248}
of $E_{8}$ breaks in
\beq
{\bf 248}=({\bf 1},{\bf 3})\oplus({\bf 56},{\bf 2})
\oplus({\bf 133},{\bf 1}).
\eeq
We are left with calculating the index theorem only for the singlet
representation, the fundamental and the adjoint representations of $SU(2)$.
Let us begin with the gaugino kinetic operator: it
contains neither $\O_{+}$ nor $\O_{-}$
but, after a simple conformal rescaling, has the form:
\beq
\g^{\m}\left(D_{\m}-\frac{1}{12}H_{\m\n\r}\g^{\n\r}\right)\chi^{I}=0.
\eeq
The volume contribution to the Dirac index theorem is then:
\beqa
{\rm ind}\Ds_{V}({\rm volume})&=&
\frac{2t+1}{192\pi^{2}}\int_{M}\tr(\tilde{R}(\om+H)
\wedge\tilde{R}(\om+H))-\cr
&-&\frac{1}{8\pi^{2}}\int_{M}\Tr_{V}(F\wedge F).
\eeqa
In presence of the CHS soliton, the
generalised curvature 2-form has components:
\beqa
\tilde{R}^{i0}&=&(\Phi^{\prime}+r\Phi^{\prime\prime})\de r\wedge\sb^{i}+
\eps^{i}{}_{jk}\sb^{j}\wedge\sb^{k}(r^{2}\Phi^{\prime 2}+r\Phi^{\prime}),\cr
\tilde{R}^{jk}&=&(2r^{2}\Phi^{\prime 2}+2r\Phi^{\prime})\sb^{k}
\wedge\sb^{j}-
\eps^{jk}{}_{i}(\Phi^{\prime}+r\Phi^{\prime\prime})\de r\wedge\sb^{i}.
\eeqa
A straightforward calculation yields:
\beq\label{volume}
{\rm ind}\Ds_{V}({\rm volume})=\frac{2t+1}{12}+\frac{2}{3}t(t+1)(2t+1);
\eeq
as far as the boundary corrections are concerned, there are two
contributions from the two boundaries of the single semi-wormhole solution,
the $S^{3}$ in the origin and the $S^{3}$ at infinity; the second
fundamental form has only components normal to the boundaries, that is
$\theta^{i}{}_{0}=\om^{i}{}_{0}\neq 0$. In this case since $\om^{i}{}_{0}$
vanishes in the origin and the curvature vanishes at infinity, there is no
local boundary correction to the index theorem.
Therefore we are left with computing the
non-local Atiyah-Patodi-Singer $\eta$--invariant correction
defined by \cite{APS}
\beq
\lim_{s\to 0}\eta(\Ds_{V},\p M,s)=
\lim_{s\to 0}\sum_{\l\neq 0}|\l|^{-s}{\rm sign}\l,
\eeq
and the dimension $h(\Ds_{V},\p M)$ of the space of the harmonic functions
of the operator $\Ds_{V}^{2}$ on the boundary, that is the number of zero
eigenvalues of the Dirac operator on the boundary.
The calculation of the $\eta$--invariant on the
$S^{3}$ at infinity is trivial, since the torsion vanishes and from the
calculation of Hitchin \cite{HIT} we get $\eta(S^{3}_{\infty})=0$,
while the same calculation is much less trivial in the origin where the
torsion contributes.
For the Dirac singlet equation one has simply:
\beq
\Ds_{\underline{1}}\chi=\sb^{\m}\left(\p_{\m}+\frac{1}{4}\om_{\m}
{}^{ab}\g_{ab}-\frac{1}{12}H_{\m}{}^{ab}\g_{ab}\right)\chi=0,
\eeq
and the Dirac operator on the $S^{3}$ in the origin becomes:
\beq
-i\Ds_{\underline{1}}=2\left(\begin{array}{cc}
K_{3}&K_{-}\\
K_{+}&-K_{3}\end{array}\right)+\uno.
\eeq
where the operators $K_{3}, K_{\pm}$ are defined in Appendix A.
The details of this calculation also can be found in Appendix A
and the final
result for the singlet is
\beq\label{etasing}
\eta(\Ds_{\underline{1}},\p M)=\lim_{s\to 0}
\sum_{k\in{\bf N}}^{\infty}\frac{2(l+1)}{(l+1)^{s}}=2\z(-1,1)=-\frac{1}{6};
\eeq
taking into account that the 3-sphere in the origin has
orientation opposite to the 3-sphere at infinity which we choose positively
oriented, and taking the sum of
(\ref{volume}) and (\ref{etasing})
we immediately get:
\beq
{\rm ind}\Ds_{\underline{1}}(M,\p M)=\left.\frac{2t+1}{12}\right|_{t=0}
+\half\cdot-\frac{1}{6}=0.
\eeq
Thus there are no normalizable zero-modes of the Dirac singlet equation,
just as in any manifold
without torsion which has a (anti)self-dual curvature two-form,
like flat space and gravitational instantons \cite{EGH}.
Let us now study
the Dirac index for the other two representations of $SU(2)$
First of all the gaugino kinetic operator on the boundary has the
form: \beq
-i\Ds\chi^{I}=-i\sb^{i}\left[\left(\p_{i}+\frac{1}{4}\om_{i}{}^{ab}\g_{ab}
-\frac{1}{12}H_{i}{}^{ab}\g_{ab}\right)\d^{I}_{K}+\half i
f^{I}{}_{JK}A^{J}_{i}
\right]\chi^{K}=0,
\eeq
where $i,a,b=1,2,3$ and where $f^{I}{}_{JK}$ are the structure constants of
the gauge group.
First we calculate the APS invariant for the fundamental representation
of $SU(2)$: since $\lim_{r\to 0}A^{I}=2\sb^{I}$, after some manipulations
we get \cite{BFRM,POPE}
\beq
-i\Ds_{\underline{2}}|_{\p M}=\left(\begin{array}{cc}
-i\Ds_{\underline{1}} &0\\
0&-i\Ds_{\underline{1}}\end{array}\right)+
\left(\begin{array}{cc}\t_{3}& \t_{-}\\
\t_{+}& -\t_{3}\end{array}\right).
\eeq
where $\t_{i}$ are the Pauli matrices and $\t_{\pm}=\t_{1}\pm i\t_{2}$.
The calculations are shown in Appendix B and it turns out that
\beq\label{etatot}
\eta(\Ds_{\underline{2}},\p M)=-\frac{1}{3};
\eeq
summing the contributions of (\ref{etatot}) and
(\ref{volume}) we get:
\beqa
{\rm ind}(\Ds_{\underline{2}},M,\p M)&=&\frac{2t+1}{12}+
\frac{2}{3}t(t+1)(2t+1)+\half\eta(\Ds_{\underline{2}},\p M)=\cr
&=&\frac{1}{6}+1-\frac{1}{6}=1,
\eeqa
that is we have only one zero-mode of the Dirac operator belonging to the
fundamental representation of $SU(2)$, the same result than in the flat
space 't Hooft instanton case, arising here from non-trivial
cancellations between the gravitational term and the non-local boundary
corrections.
As for the triplet case, we follow the same steps and
we finally find that
and the Dirac index for the $SU(2)$ triplet becomes:
\beq
{\rm ind}\Ds_{\underline{3}}(M,\p M)=\frac{1}{4}+4-\frac{1}{4}=4.
\eeq
The calculations for this case can be found in Appendix C.
As in the previous case, we find a non trivial cancellation between the
gravitational part and the $\eta$--correction which leaves us with
the same result as for flat space.

If, before making the conformal rescaling on
the gaugino, we had used explicitly the form of $H_{\m\n\r}$, we would have
found that the torsion contribution is comparable to the rescaled term
so that it could be rescaled away together with all the other terms
containing the dilaton. In
this case we should apply to the gaugino the usual index theorem
without torsion. As a simple exercise let us verify that our
previous results are
consistent with this picture.
The curvature 2-form now has the form:
\beq
R^{i}{}_{0}=(\Phi^{\prime}+r\Phi^{\prime\prime})\de r\wedge\sb^{i},\quad
R^{j}{}_{k}=(r^{2}\Phi^{\prime 2}+2r\Phi^{\prime})\sb^{j}\wedge\sb^{i},
\eeq
and $\int_{M}\tr(R(\om)\wedge R(\om))=0$,
so that the bulk contribution to the index theorem comes entirely from
the gauge fields; as for the $\eta$--invariant, it must be calculated
on two $S^{3}$ and for both of them it is zero.
The indices for Dirac fermions are then:
\beq
{\rm ind}\Ds_{\underline{1}}(M,\p M)=0,\quad
{\rm ind}\Ds_{\underline{2}}(M,\p M)=1,\quad
{\rm ind}\Ds_{\underline{3}}(M,\p M)=4,
\eeq
exactly the result we obtained through a separate treatment of the torsion
term in the equations of motion.

As for the dilatino, the calculation is trivial, since neither torsion
nor gauge fields couple to its kinetic operator.
Using our previous results, we obtain immediately that
\beq
{\rm ind}(\Ds_{\l},M,\p M)=0;
\eeq
in principle one should be careful in making singular rescalings of the
fermionic fields to get rid of the dilatonic terms in the
kinetic operators and should wonder whether the number of zero-modes
of the rescaled fermions is the same of the original ones.
As a check we have verified that the original dilatino kinetic
operator does not admit zero-modes; using the fact that possible
dilatino zero-modes should have the form
\beq
\d\l_{+}=-\frac{1}{4}\left(\g^{\m}\p_{\m}\Phi-\frac{1}{6}H_{\m\n\r}
\g^{\m\n\r}\right)\eps_{-}=-\half\g^{\m}\p_{\m}\Phi\eps_{-},
\eeq
and putting this field configuration in the equation of motion,
if we make the plausible Ansatz $\eps_{-}=e^{p\Phi}\eta$ with constant
$\eta$, we find a non-normalizable solution for $p=\frac{5}{2}$.
We therefore find no zero-modes for the
dilatino qualitatively
justifying the singular conformal rescaling we applied before.
Regarding the gravitino field, using
the explicit form of the metric and computing the spin
connection appearing in the covariant derivative,
we obtain an alternative form of the
equation of motion for the gravitino:
\beq
\g^{\m\n\r}\hat{D}_{\n}\psi_{\r}^{-}+
2\p^{\m}\Phi\g^{\n}\psi_{\n}^{-}
-\half H^{\m\n\r}\g_{\n}\psi_{\r}^{-}=0,
\eeq
where the covariant derivative $\hat{D}_{\m}$ is evaluated with respect
to the metric $\hat{g}_{\m\n}=e^{-2\Phi}\d_{\m\n}$.
Now we
can make a gauge choice on the gravitino which does not affect the physics
of the problem and we choose a condition of
$\g$-tracelessness; with this information we can immediately throw away
the second term in the equation of motion and we are left with just a
torsion term.
Working out the index theorem
for the Rarita-Schwinger
operator with torsion has been proved so far very
hard to complete and we are
still working on it;
moreover the coupling of $H_{\m\n\r}$ to the gravitino does not
seem to be compatible with its interpretation as the
antisymmetric part of the affine connection so that
it is not clear how to generalise the index theorem to cope with this
specific case.
We would however like
to show that if we use the explicit
form of the torsion in this solution, we can absorb the torsion term
through a Weyl rescaling of the gravitino field:
in fact,
\beq
H^{\m\n\r}\g_{\n}\psi_{\r}^{-}=-\sqrt{G}\eps^{\m\n\r\s}\p_{\s}\Phi\g_{\n}
\psi_{\r}^{-}=\G^{\m\n\r}\p_{\n}\Phi\psi_{\r}^{-},
\eeq
and defining a new gravitino $\hat{\psi}_{\m}=e^{-\Phi/2}\psi_{\m}$ we
obtain a new equation of motion:
\beq
\g^{\m\n\r}\hat{D}_{\n}\hat{\psi}_{\r}^{-}=0,
\eeq
and we can work with the
usual index theorem.
It remains to verify whether this procedure is correct, through an extension
of the Rarita-Schwinger index formula to the case of coupling to torsion.
In absence of torsion, the index of the Rarita-Schwinger operator
is given by the formula:
\beqa
{\rm ind}D_{RS}(M,\p M)&=&\!-\frac{21}{192\pi^{2}}
\left[\int_{M}\tr\ R(\om)
\wedge R(\om)-\int_{\p M}
\tr\ \theta(\om)\wedge R(\om)\right]\nonumber\\
&-&\half[\eta_{RS}(\p M)+h_{RS}(\p M)].
\eeqa
In this case we must be careful in computing curvature 2-forms and
fundamental forms with the metric $\hat{g}_{\m\n}$.
One can check that Rarita-Schwinger $\eta$--invariant on $S^{3}$ is zero so
that there are no normalizable solution to the Rarita-Schwinger equation.
We have also explicitly checked that, by making the rather general ansatz
$\eps_{-}=e^{p\Phi}\eta$ with $p$ and $\eta$ constants, the original
Rarita-Schwinger operator has no zero-modes, enforcing our
qualitative results.
{}From the property of factorisation of the generalised Dirac determinant, we
can infer that the fermionic zero-modes could in principle have three
different forms:
\beq
\left(\begin{array}{c}
\hat{\psi}_{\m}^{(0)}\\
\hat{\chi}^{a\prime}\\
\hat{\l}^{\prime}\end{array}\right), \quad\quad
\left(\begin{array}{c}
0\\
\hat{\chi}^{a}_{(0)}\\
\hat{\l}^{\prime\prime}\end{array}\right),\quad\quad
\left(\begin{array}{c}
0\\
0\\
\hat{\l}^{(0)}\end{array}\right),
\eeq
where the index ``0" stands for the zero-modes, while the other entries are
relative to the solution of the
fermionic equations of motion in presence of
zero-modes; from our analysis we have found that only solutions of the
second kind can exist and that there is no space for gravitino condensates.
The four gaugino triplet zero-modes are expected to correspond
to the breaking of
two supersymmetries and of two superconformal symmetries of the theory as
it happens in the usual flat Super-Yang-Mills case and to have the form
\beq
\hat{\chi}^{(0)}=-\frac{1}{4}e^{-\Phi}F_{\m\n}\g^{\m\n}\eta,
\eeq
where $\eta=\eta_{0}+\r^{-1}x^{\m}\g_{\m}\epsb_{0}$ and
$\eta_{0},\epsb_{0}$ are constant spinors.
As for the dilatino and the gravitino, from a more geometrical point
of view we would not expect to find any zero-modes, since zero-modes,
at least in the single instanton case, are
related to broken symmetries and in this case they are just supersymmetry
and superconformal symmetry leaving no space for other zero-modes to exist.
We have also found explicitly that associated to the two
``supersymmetric" gaugino zero-modes the dilatino has a non trivial
normalizable configuration $\hat{\l}_{+}=-\half\g^{\m}\p_{\m}\Phi\eta_{0}$
which satisfies the equations of motion for the dilatino in presence of the
gaugino zero-modes, while for the other two zero-modes the dilatino field
is $\hat{\l}_{+}=-\half\g^{\m}\p_{\m}\Phi(\r^{-1}x^{\l}\g_{\l}\epsb_{0})$
which is not normalizable.
\section{Conclusions}
We have set up a machinery which allows the calculation of the
Dirac index theorem both in presence and in absence of torsion; its
usefulness stems out from the fact that torsion appears naturally in all
superstring inspired supergravity theories and it couples to fermions.
It remains to examine the gravitino case which does not couple to
torsion according
to naive expectations; it has been proved that the Rarita-Schwinger
index is not affected by the presence of torsion
\cite{URRU}, but unfortunately
the kind of coupling taken into account, although consistent,
is not the one that appears in supergravity Lagrangians.
At this point it would be interesting
to apply the above method to the computation of
topological invariants for string solutions related by Buscher's duality
\cite{BUSCH}. If T-duality is sensibly performed, it relates models with
torsion and conformally flat metrics to models without torsion and
non-trivial metrics;
it has been shown, for example, that the dual of CHS is a kind of black
holes \cite{AHN}.
In general, however, if the Killing vector under which we perform
the duality has some zeros the dual solution is singular so that
the computation
may require a more accurate analysis.
When a conformal field theory approach is applicable one may find a
(non-local) correspondence among vertex operators so that the number
of zero modes should agree between T-dual solutions.
Anyway our method has a wider applicability being independent
of the possibility
of a more formal CFT approach and could provide a new insight on the
behaviour of all topological invariants under these duality
transformations.
We have restricted our attention to $D=4$, but it is straightforward in
principle to extend the analysis to higher dimensional manifolds with
torsion. In the compact case this has been done by Rohm and Witten
\cite{ROHM}; recently however non-compact manifolds seem to receive
new interest \cite{POLCH}.

{\Large{\bf Acknowledgements}}\\
I would like to thank M. Bianchi and my Ph.D. advisor F. Fucito for
suggesting me the problem and for their constant support and
C. Angelantonj for clearing me one point in the present article.
\newpage

\appendix{{\Huge Appendix A}}
\\ \noindent
Let us define the right-invariant one--forms:
\beq
\bar\sigma_i = {1\over {r^2}} \bar\eta_{i\m\n} x^\mu \de x^\nu,
\eeq
or in terms of Euler angles:
\beqa
\bar\sigma_x&=&{1\over 2} (\sin{\phi}\ \de\theta -
 \cos{\phi} \sin{\theta}\ \de\psi),\nonumber\\
\bar\sigma_y&=&-{1\over 2} ( \cos{\phi}\ \de\theta +
 \sin{\phi} \sin{\theta}\ \de\psi),\\
\bar\sigma_z&=&{1\over 2} ( \de\phi + \cos{\theta}\ \de\psi)\nonumber.
\eeqa
Their dual vector fields are
\beq
K_i = - {i\over 2} \bar\eta_i{}^{\m\n} x_\mu \p_\nu,
\eeq
which satisfy the $SU(2)$ algebra: $[K_i,K_j] =
-i \varepsilon_{ij}^k K_k$.
The Dirac singlet equation on a generic boundary is simply:
\beq
\Ds_{\underline{1}}\chi=\sb^{\m}\left(\p_{\m}+\frac{1}{4}\om_{\m}
{}^{ab}\g_{ab}-\frac{1}{12}H_{\m}{}^{ab}\g_{ab}\right)\chi=0,
\eeq
and the Dirac operator on the $S^{3}$ in the origin becomes:
\beq
-i\Ds_{\underline{1}}=2\left(\begin{array}{cc}
K_{3}&K_{-}\\
K_{+}&-K_{3}\end{array}\right)+\uno,
\eeq
where $K_{\pm}=K_{1}\pm iK_{2}$.
To calculate the eigenvalues of this operator let us act with it upon
a state $\Psi$ whose entries are the $SU(2)$ rotation matrices
$D_{n,m}^{l}(\theta,\phi,\psi)$
\beq
\Psi=\left(\begin{array}{c}
D_{n,m-1}^{l}\\
D_{n,m}^{l}\end{array}\right).
\eeq
The Dirac operator becomes:
\beq
-i\Ds_{\underline{1}}\Psi=2\left(\begin{array}{cc}
m-\half& a\\
a&-m+\half\end{array}
\right)\Psi,
\eeq
where $a=\sqrt{(l+m)(l-m+1)}$. The eigenvalues of this matrix are
$2l+1, -2l-1$ with multiplicities $d=2l(2l+1)$
since we restrict $m$ to the range $-l+1\leq m\leq l$.
Otherwise some components of the eigenvector
would lose their meaning and
we have to study these limiting cases separately \cite{POPE}:
\begin{enumerate}
\item $m=-l$. In this case the first component of the eigenvector must
be set to zero and we have a single eigenvector with eigenvalue $2l+1$
and multiplicity $d=2l+1$ equal to the degeneracy of the quantum number
$n$;
\item $m=l+1$. The last component of the eigenvector must be set to zero
and again we have one eigenvector with eigenvalue $2l+1$ and multiplicity
$d=2l+1$.
\end{enumerate}
The contribution to the $\eta$--invariant from the general cases is
identically zero, so we have just to evaluate the contribution coming from the
limiting cases:
\beq
\eta(\Ds_{\underline{1}},\p M)=\lim_{s\to 0}
\sum_{l=0}^{\infty}\frac{2(2l+1)}{(2l+1)^{s}},
\eeq
where the sum runs over integers and half-integers.
Turning to a sum over integers we can reexpress this sum in term of
generalised Riemann $\z$-functions:
\beq
\eta(\Ds_{\underline{1}},\p M)=2\zeta(-1,1)=-\frac{1}{6}.
\eeq
\appendix{{\Huge Appendix B}}
\\ \noindent
In this appendix we will explicitly perform the calculation of the
 $\eta$--invariant for a gaugino in the fundamental representation
of $SU(2)$.
The Dirac equation on the boundary is
\beq
-i\Ds_{\underline{2}}|_{\p M}=\left(\begin{array}{cc}
-i\Ds_{\underline{1}} &0\\
0&-i\Ds_{\underline{1}}\end{array}\right)+
\left(\begin{array}{cc}\t_{3}& \t_{-}\\
\t_{+}& -\t_{3}\end{array}\right).
\eeq
We now have to look for the eigenvalue spectrum
of the matrix ${\cal M}$
defined by
\beq
{\cal M}=2\left(\begin{array}{crcc}
K_{3}+1&K_{-}&0&0\\
K_{+}&-K_{3}&1&0\\
0&1&K_{3}&K_{-}\\
0&0&K_{+}&-K_{3}+1\end{array}\right)
+\uno.
\eeq
The form of this matrix suggests us that we can solve the eigenvalue
equation ${\cal M}\Psi=\l\Psi$ by
expanding $\Psi$ in terms of the $SU(2)$ rotation matrices
$D^{l}_{n,m}(\theta,\phi,\psi)$ which are defined in the appendix. We are
left with a finite dimensional eigenvalue problem for a $4\times 4$ matrix;
for generic, given values of $l,n,m$ (the operator ${\cal M}$
does not act over
the index $n$) there are four eigenvectors that may be written as:
\beq\label{eigenvect}
\Psi_{n,m}^{(i)l}=\left(\begin{array}{c}
c_{1}^{i}D^{l}_{n,m-1}\\
c_{2}^{i}D^{l}_{n,m}\\
c_{3}^{i}D^{l}_{n,m}\\
c_{4}^{i}D^{l}_{n,m+1}\end{array}\right),\;\;\;i=1,2,3,4.
\eeq
The explicit expression of the
coefficients $c_{k}^{(i)}$ is irrelevant to our
problem.
Since the allowed values of
$n,m$ in the (\ref{eigenvect}) are
restricted to be $|m|\leq l-1$ and $|n|\leq l$ if $l\neq 0$,
let us again study the
limiting cases separately. The four eigenvalues corresponding to the
eigenvectors
(\ref{eigenvect})  are
\beqa
& &\l^{(1)l}_{n,m}=2l+2,\;\;\;\l^{(2)l}_{n,m}=2l,\\
& &\l^{(3)l}_{n,m}=-2l\;\;\;\l^{(4)l}_{n,m}=-2l-2;
\eeqa
they have the same multiplicity $d$, given by the product of the
degeneracies of the $n$ and $m$ quantum numbers:
$d=d_{m}d_{n}=(2l+1)(2l-1)$.
Since these eigenvalues are symmetrically distributed around zero, they
do not contribute to the $\eta$--invariant. Let us now investigate the
contribution of the limiting cases in which either $l=0$ or
$m$ violates the condition $|m|\leq l-1$. As far as the special case
$l=0$ is concerned, the matrix ${\cal M}$ becomes simply:
\beq
{\cal M}=\left(\begin{array}{crrc}1&0&0&0\\
0&-1&2&0\\
0&2&-1&0\\
0&0&0&1\end{array}\right)+\uno,
\eeq
whose eigenvectors are constant and eigenvalues 2,-2 with multiplicities
$d=3,1$; the contribution from this case is very easily calculated to be
\beq\label{etaspec}
\lim_{s\to 0}\eta(\Ds_{\underline{2}},\p M,s)_{l=0}
=\lim_{s\to 0}\left(\frac{3}{2^{s}}-
\frac{1}{2^{s}}\right)=2.
\eeq
Finally let us turn to the other limiting cases; they are dealt
with by setting
to zero the components of (\ref{eigenvect}) that lose meaning when
$m$ takes its limiting values $m=\pm l$ or $m=\pm(l+1)$.
The cases are the following:
\begin{enumerate}
\item $m=l+1$. Only the first component in (\ref{eigenvect}) is non-zero;
there is one eigenvector with eigenvalue $2l+2$ and multiplicity $d=2l+1$;
\item $m=-l-1$. The last component in (\ref{eigenvect}) is non-zero: there
is again one eigenvector with eigenvalue $2l+2$ and multiplicity $d=2l+1$;
\item $m=l$ The first three components in (\ref{eigenvect})
are non vanishing; there are 3 eigenvectors $\Psi^{(k)l}_{l,n}$,
$k=1,2,3$ with eigenvalues $2l+2,2l,-2l-2$ with multiplicity $d=2l+1$;
\item $m=-l$ The last three components in (\ref{eigenvect})
are non vanishing; there are 3 eigenvectors $\Psi^{(k)l}_{-l,n}$,
$k=1,2,3$ with eigenvalues $2l+2,2l,-2l-2$ with multiplicity $d=2l+1$.
\end{enumerate}
The contribution from these four cases is then
\beqa\label{etalimit}
& &\lim_{s\to 0}\eta(\Ds_{\underline{2}},\p M,s)|_{l\neq 0}=
\sum_{l=1/2}^{\infty}\left(\frac{4(2l+1)}{(2l+2)^{s}}+\frac{2(2l+1)}
{(2l)^{s}}-\frac{2(2l+1)}{(2l+2)^{s}}\right)=\cr
& &=2\z(-1,3)-2\z(0,3)+2\z(-1,1)+2\z(0,1)=-\frac{7}{3};
\eeqa
adding (\ref{etalimit}) to (\ref{etaspec}) and since there are no zero
eigenvalues of $\Ds_{\underline{2}}$ on the boundary we obtain that
\beq
\eta(\Ds_{\underline{2}},\p M)=\eta(\Ds_{\underline{2}},\p M)_{l=0}+
\eta(\Ds_{\underline{2}},\p M)_{l\neq 0}=2-\frac{7}{3}=-\frac{1}{3}.
\eeq

\appendix{{\Huge Appendix C}}

\noindent
In this appendix we want to calculate explicitly the value of the $\eta$--
invariant for the adjoint representation of $SU(2)$: following the same
steps as in appendix B and using for simplicity a form for
the generators in the adjoint representation which has $T_{3}$ diagonal,
that is:
\beqa
T_{1}&=&\frac{1}{\sqrt{2}}\left(\begin{array}{ccc}
0&-1&0\\
-1&0&1\\
0&1&0\end{array}\right),\;\;\;T_{2}=\frac{1}{\sqrt{2}}
\left(\begin{array}{ccc}
0&i&0\\
-i&0&-i\\
0&i&0\end{array}\right),\cr
T_{3}&=&\left(\begin{array}{ccc}
1&0&0\\
0&0&0\\
0&0&-1
\end{array}\right),
\eeqa
the Dirac operator will take the form:
\beq
-i\Ds_{\underline{3}}|_{\p M}=
\left(\begin{array}{ccc}
-i\Ds_{\underline{1}}&0&0\\
0&-i\Ds_{\underline{1}}&0\\
0&0&-i\Ds_{\underline{1}}\end{array}\right)+
\left(\begin{array}{ccc}
2\t_{3}&-\sqrt{2}\t_{-}&0\\
-\sqrt{2}\t_{+}&0&\sqrt{2}\t_{-}\\
0&\sqrt{2}\t_{+}&-2\t_{3}\end{array}
\right).
\eeq
Again, for generic, given, values of $l$, $m$ and $n$, with $l\neq 0$,
$|n|\leq l$ and $-l+1\leq m\leq l-2$ there are six different eigenvectors
whose form is given by:
\beq\label{eigen3}
\Psi_{nm}^{(i)l}=\left(\begin{array}{c}
c_{1}^{i}D^{l}_{n,m-1}\\
c_{2}^{i}D^{l}_{n,m}\\
c_{3}^{i}D^{l}_{n,m}\\
c_{4}^{i}D^{l}_{n,m+1}\\
c_{5}^{i}D^{l}_{n,m+1}\\
c_{6}^{i}D^{l}_{n,m+2}
\end{array}\right),\;\;\;i=1,\ldots,6.
\eeq
The corresponding eigenvalues are:
\beqa
& &\l^{(1)l}_{n,m}=2l+3,\;\;\;\l^{(2)l}_{n,m}=2l+1,\\
& &\l^{(3)l}_{n,m}=2l-1,\;\;\;\l^{(4)l}_{n,m}=-2l-3,\\
& &\l^{(5)l}_{n,m}=-2l-1,\;\;\;\l^{(6)l}_{n,m}=-2l+1;
\eeqa
since they are symmetrically distributed around zero and have the
same multiplicity, again
they do not contribute to the $\eta$--invariant. Let us study the remaining
seven limiting cases:
\begin{enumerate}
\item $l=m=n=0$. In this case the operator has two constant eigenvectors with
eigenvalues 3,--3 and multiplicities d=4,2 respectively. Their contribution to
the
$\eta$ is simply $\eta_{l=0}=2$;
\item $m=l+1$. Only the first component in (\ref{eigen3}) is non-zero; there is
one eigenvector with eigenvalue $2l+3$ and multiplicity $d=2l+1$;
\item $m=-l-2$. The last component in (\ref{eigen3}) is non-zero; we have one
eigenvector with eigenvalue $2l+3$ and multiplicity $d=2l+1$;
\item $m=l$. The last three components in (\ref{eigen3}) must be set to zero;
there are three eigenvectors $\Psi^{(k)l}_{nl}$, with $k=1,2,3$ whose
eigenvalues are $2l+3, 2l+1, -2l-3$, with multiplicity $d=2l+1$;
\item $m=-l-1$. This case is analogous to the previous one, except that the
first three components in (\ref{eigen3}) are zero. Three eigenvectors with
eigenvalues $2l+3, 2l+1, -2l-3$, and multiplicity $d=2l+1$.
\item $m=l-1$ and $m=-l$. In both cases we have five eigenvectors with
eigenvalues $2l+3, 2l+1, 2l-1, -2l+1, -2l-3$, with multiplicity $d=2l=1$.
\end{enumerate}
Putting all these contributions together we obtain:
\beq
\eta(\Ds_{\underline{3}},\p M)=\eta_{_{l=0}}
+\eta_{_{limit}}=2-\frac{5}{2}=-\half,
\eeq

\end{document}